\documentclass{PoS}
\usepackage{amsmath,amssymb}
\usepackage{array, booktabs, cite}

\title{Renormalization on the fuzzy sphere}

\ShortTitle{Renormalization on the fuzzy sphere}

\author{
\speaker{Kohta Hatakeyama}$^{1,2}$, Asato Tsuchiya$^{1,2}$ and Kazushi Yamashiro$^{1}$\\
        $^1$Department of Physics, Shizuoka University, 
        836 Ohya, Suruga-ku, Shizuoka 422-8529, Japan\\
        $^2$Graduate School of Science and Technology, Shizuoka University, 
        3-5-1 Johoku, Naka-ku, \phantom{x}Hamamatsu 432-8011, Japan\\
        E-mail: \email{hatakeyama.kohta.15@shizuoka.ac.jp}, 
        \email{tsuchiya.asato@shizuoka.ac.jp}, 
        \email{yamashiro.kazushi.17@shizuoka.ac.jp}
        }

\abstract{
We study renormalization on the fuzzy sphere, 
which is a typical example of non-commutative spaces. 
We numerically simulate a scalar field theory on the fuzzy sphere, 
which is described by a Hermitian matrix model. 
We define correlation functions by using the Berezin symbol and 
show that they are made independent of the matrix size, which plays a role of a UV cutoff, 
by tuning one parameter of the theory.
We also find that the theories on the phase boundary are universal.
They behave as a conformal field theory at short distances, 
while they universally differ from it at long distances due to the UV/IR mixing.
}

\FullConference{The 36th Annual International Symposium on Lattice Field Theory - LATTICE2018\\
		22-28 July, 2018\\
		Michigan State University, East Lansing, Michigan, USA.}

\begin{document}

\setcounter{footnote}{0}

\section{Introduction}
\setcounter{equation}{0}
\renewcommand{\thefootnote}{\arabic{footnote}} 
It is known that field theories on non-commutative spaces are deeply involved in quantum gravity or string theory (for a review, see \cite{Douglas:2001ba}).
One of the most characteristic phenomena in field theories on non-commutative spaces
is the so-called UV/IR mixing \cite{Minwalla:1999px}, which prevents perturbative renormalization.

It is important to resolve the problem of renormalization 
in order to construct consistent quantum field theories on non-commutative spaces.
It was shown by Monte Carlo simulation in \cite{Hatakeyama:2017fao, Hatakeyama:2018qjr} 
that the 2-point and 4-point correlation functions in the disordered phase of 
a scalar field theory on the fuzzy sphere\footnote{By Monte Carlo simulation, the theory has been studied in \cite{Martin:2004un,Panero:2006bx,Panero:2006cs,
GarciaFlores:2009hf,Das:2007gm,Hatakeyama:2017fao, Hatakeyama:2018qjr}. 
The model has also been studied analytically in \cite{Kawamoto:2015qla, Vaidya:2003ew, OConnor:2007ibg, Nair:2011ux, Polychronakos:2013nca, Tekel:2013vz, Saemann:2014pca, Tekel:2014bta, Tekel:2015zga}.} 
are made independent of the matrix size up to a wave function renormalization 
by tuning a parameter in the theory, 
where the matrix size is regarded as a UV cutoff\footnote{In \cite{Bietenholz:2004xs,Mejia-Diaz:2014lza}, 
a scalar field theory on the non-commutative torus was studied in a similar way.}.
This strongly suggests that the theory is non-perturbatively renormalizable in the disordered phase.

Here, we perform further study of the scalar field theory
on the fuzzy sphere by Monte Carlo simulation \cite{Hatakeyama:2018qjr}.
Multi-point correlation functions are defined by using the Berezin symbol \cite{Berezin:1974du}
in the same way as in \cite{Hatakeyama:2017fao}.
Then, the phase boundary is identified by calculating the susceptibility 
which is an order parameter for the $Z_2$ symmetry,
and the 2-point and 4-point correlation functions are calculated on the boundary. 
It is found that the correlation functions at different points on the boundary agree 
so that the theories on the boundary are universal as in ordinary field theories. 
Furthermore, we find that the behavior of the 2-point correlation functions is the same as 
that in a conformal field theory (CFT) at short distances but different from it at long distances.

\section{Scalar field theory on the fuzzy sphere}
\setcounter{equation}{0}
We study the following matrix model:
\begin{equation}
S=\frac{1}{N}\mbox{Tr}\left(-\frac{1}{2}[\hat{L}_i,\Phi]^2+\frac{\mu^2}{2}\Phi^2
+\frac{\lambda}{4}\Phi^4\right) \ ,
\label{action}
\end{equation}
where $\Phi$ is an $N\times N$ Hermitian matrix, and
$N \times N$ matrices $\hat{L}_i$ ($i=1,2,3$) are the generators of the $SU(2)$ algebra 
in the $N$-dimensional irreducible representation, 
which satisfy the commutation relation $[\hat{L}_i,\hat{L}_j]=i\epsilon_{ijk} \hat{L}_k$.
The theory possesses $Z_2$ symmetry: $\Phi \rightarrow -\Phi$.
The path-integral measure is defined by $d\Phi e^{-S}$, where
\begin{equation}
d\Phi=\prod_{i=1}^N d\Phi_{ii} \prod_{1\leq j < k\leq N}
d\mbox{Re}\Phi_{jk} d\mbox{Im}\Phi_{jk}\ .
\end{equation}

At the tree level in the $N\rightarrow \infty$ limit, which corresponds
to the so-called commutative limit,
the theory (\ref{action}) reduces to the following continuum theory
on a sphere with the radius $R$:
\begin{equation}
S_\mathrm{C}=\int \frac{R^2 d\Omega}{4\pi} \left(-\frac{1}{2R^2}({\cal L}_i\phi (\Omega))^2
+\frac{m^2}{2}\phi(\Omega)^2+\frac{g}{4}\phi(\Omega)^4\right) 
\label{continuum action}
\end{equation}
with the invariant measure on the sphere $d\Omega$ and 
the orbital angular momentum operators ${\cal L}_i$ ($i=1,2,3$). 

In the above correspondence,
the Berezin symbol $\langle\Omega|\Phi|\Omega\rangle$, which is explained in the next section,
is identified with the field $\phi(\Omega)$, and 
the parameters in (\ref{action}) correspond to those in (\ref{continuum action}) as
$
\mu^2  = R^2 m^2, \ \lambda  = R^2 g
$.
The authors of \cite{Chu:2001xi,Steinacker:2016nsc} showed
that the 1-loop effective action of (\ref{action}) differs from that of (\ref{continuum action}) 
by a finite amount.

\section{Correlation functions}
\setcounter{equation}{0}
We define correlation functions 
by introducing the Berezin symbol \cite{Berezin:1974du} constructed from
the Bloch coherent state \cite{Gazeau}.
The sphere is parametrized by the standard polar coordinates $\Omega=(\theta,\varphi)$.
The Bloch coherent state $|\Omega \rangle$
is localized around the point on the sphere, $(\theta,\varphi)$, with the width $R/\sqrt{N}$.
For an $N\times N$ matrix $M$, 
the Berezin symbol is defined by $\langle\Omega|M|\Omega\rangle$.

In the following, we denote the Berezin symbol shortly as 
$
\varphi(\Omega) = \langle\Omega | \Phi | \Omega\rangle
$.
Then, the $n$-point correlation function in the theory (\ref{action}) is defined as
\begin{equation}
\left\langle \varphi(\Omega_1)\varphi(\Omega_2)\cdots\varphi(\Omega_n)
\right\rangle
=\frac{\int d\Phi \ \varphi(\Omega_1)\varphi(\Omega_2)\cdots\varphi(\Omega_n) 
 e^{-S} }{ \int d\Phi \  e^{-S}} \ .
\label{n-point function}
\end{equation}
This correlation function is a counterpart of 
$\langle \phi(\Omega_1)\phi(\Omega_2)\cdots\phi(\Omega_n)\rangle$ 
in the theory (\ref{continuum action}).

We assume the matrix $\Phi$ in (\ref{action}) to be renormalized as
$
\Phi=\sqrt{Z}\Phi_\mathrm{r} 
$
with the renormalized matrix $\Phi_\mathrm{r}$.
Then, the renormalized Berezin symbol is defined by
$
\varphi(\Omega)=\sqrt{Z}\varphi_\mathrm{r}(\Omega)
$,
and the renormalized $n$-point correlation function is defined by
\begin{equation}
\left\langle \varphi(\Omega_1)\varphi(\Omega_2)\cdots\varphi(\Omega_n)
\right\rangle
=Z^{\frac{n}{2}}
\left\langle \varphi_\mathrm{r}(\Omega_1)\varphi_\mathrm{r}(\Omega_2)\cdots\varphi_\mathrm{r}(\Omega_n)
\right\rangle \ .
\end{equation}

\begin{figure}[t]
\centering
\includegraphics[width=5cm]{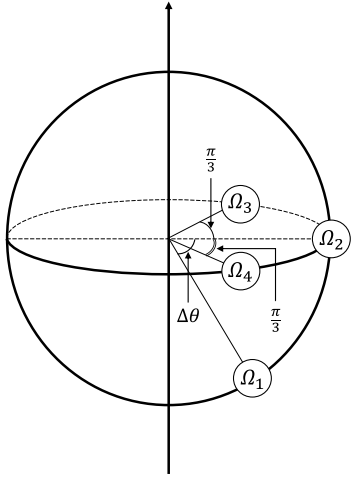}
\caption{Four points on the sphere chosen for the correlation functions.}
\label{sphere}
\end{figure}

In the following, we calculate the following correlation functions:
\begin{align}
&\mbox{2-point function:} \; \left\langle \varphi(\Omega_\rho)\varphi(\Omega_\sigma)
\right\rangle_\mathrm{c} \; (1\leq \rho < \sigma \leq 4) \ , \nonumber\\
&\mbox{4-point function:} \; \left\langle \varphi(\Omega_1)\varphi(\Omega_2)
\varphi(\Omega_3)\varphi(\Omega_4) \right\rangle_\mathrm{c} \ ,
\label{correlation functions}
\end{align}
where c stands for the connected part.

We pick up four points $\Omega_\rho=(\theta_\rho,\varphi_\rho)$ ($\rho=1,\ldots,4$) on the sphere as follows (see Fig. \ref{sphere}):
\begin{equation}
\Omega_1 = \left(\frac{\pi}{2} + \Delta \theta, \ 0\right)  \ , \quad
\Omega_2 = \left(\frac{\pi}{2} , \ 0\right)  \ , \quad
\Omega_3 = \left(\frac{\pi}{2} , \ \frac{\pi}{3}\right) \ , \quad
\Omega_4 = \left(\frac{\pi}{2} , \ \frac{5\pi}{3} \right) \ .
\label{angles on the sphere}
\end{equation}

We apply the hybrid Monte Carlo method to our simulation of the theory.

\subsection{Critical behavior of correlation functions}
\begin{figure}[t]
\centering
\includegraphics[width=12cm]{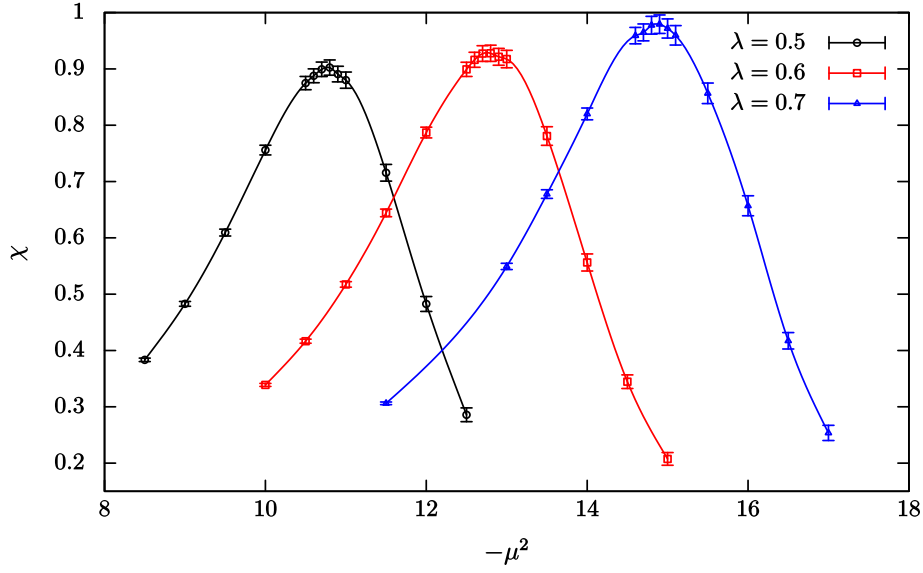}
\caption{We plot the susceptibility $\chi$ at $N=24$ against $-\mu^2$. 
The circles, the squares, and the triangles represent the data for $\lambda=0.5, 0.6, 0.7$, respectively. There are the peaks of $\chi$ for $\lambda=0.5,0.6,0.7$ around $\mu^2=-10.8, -12.8, -14.8$, respectively.}
\label{chi}
\end{figure}

We calculate the 2-point and 4-point correlation functions 
on the phase boundary at $N=24$.
To determine the phase boundary, 
we calculate the susceptibility $\chi \equiv \left\langle \left(\mbox{Tr}\Phi/N \right)^2\right\rangle -\left\langle \left|\mbox{Tr}\Phi \right|/N \right\rangle^2$ 
that is an order parameter for the $Z_2$ symmetry which is a symmetry under $\Phi \to -\Phi$.
In Fig.\ref{chi}, $\chi$ is plotted against $-\mu^2$ for $\lambda=0.5,0.6,0.7$. 
We find that the peak of $\chi$ for $\lambda=0.5,0.6,0.7$ exists around
$\mu^2=-10.8, -12.8, -14.8$, respectively.
The theory is in the unbroken phase in the left side of the peak, 
while it is in the broken phase in the right side of the peak.

We calculate 2-point and 4-point correlation functions at various $\mu^2$ around the above values 
for each $\lambda$, and find that the correlation functions for $(\mu^2, \lambda)=(-10.801, 0.5), (-12.810, 0.6)$, $(-14.925, 0.7)$ agree after performing wave function renormalization.
In the following, we show that this is indeed in the case.

For later convenience, 
we introduce a stereographic projection defined by
$
z = R\tan\frac{\theta}{2} e^{i\varphi} 
$,
which maps a sphere with the radius $R$ to the complex plane.
Here, $R$ is fixed at $1$ without loss of generality.
We calculate 
$
\langle \varphi(z_m) \varphi(1) \rangle_\mathrm{c}  
$
and 
$
\langle \varphi(z_m) \varphi(1) 
\varphi(e^{i\frac{\pi}{3}})\varphi(e^{i\frac{5\pi}{3}})\rangle_\mathrm{c} 
$,
where 
$
z_m = \tan \left[\frac{1}{2} \left(\frac{\pi}{2} + 0.1m \right) \right]
$
with $m=1,\ldots,15$.

\begin{figure}[t]
\centering
\includegraphics[width=11cm]{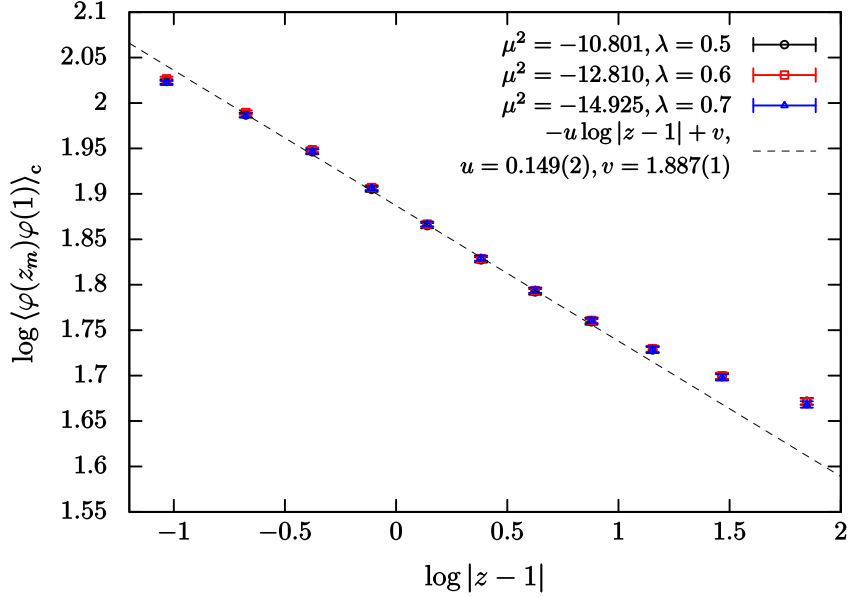}
\caption{$\log \langle \varphi(z_m) \varphi(1) \rangle_\mathrm{c}$ at $N=24$ is plotted against $\log |z-1|$. The circles, the squares, and the triangles represent the data for $\lambda=0.5, 0.6, 0.7$, respectively. The data for $\lambda=0.6, 0.7$ are simultaneously shifted by $\alpha_{0.6 \to 0.5}=-0.015(1)$ and $\alpha_{0.7 \to 0.5}=-0.056(1)$, respectively, in the vertical direction so that they agree with the data for $\lambda=0.5$. The dashed line is a fit of seven data points (from the second point to the eighth point) of $\log \langle \varphi(z_m) \varphi(1) \rangle_\mathrm{c}$ at $\lambda=0.5$ to $-u\log |z-1|+v$ with $u=0.149(2)$ and $v=1.887(1)$.}
\label{2pt_after_renormalization_oncritical}
\end{figure}

\begin{figure}[t]
\centering
\includegraphics[width=11cm]{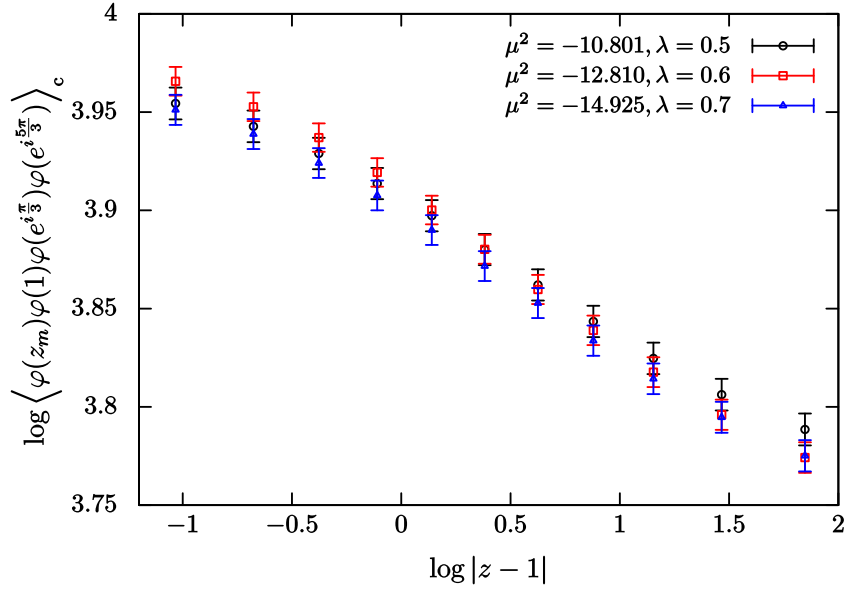}
\caption{$\log \langle \varphi(z_m) \varphi(1) \varphi(e^{i\frac{\pi}{3}})\varphi(e^{i\frac{5\pi}{3}})\rangle_\mathrm{c}$ at $N=24$ is plotted against  $\log |z-1|$.  
The circles, the squares, and the triangles represent the data for $\lambda=0.5, 0.6, 0.7$, respectively. The data for $\lambda=0.6, 0.7$ are simultaneously shifted by $2\alpha_{0.6 \to 0.5}$ and $2\alpha_{0.7 \to 0.5}$, respectively, in the vertical direction so that they agree with the data for $\lambda=0.5$.}
\label{4pt_after_renormalization_oncritical}
\end{figure}

In Fig.\ref{2pt_after_renormalization_oncritical},
we plot $\log \langle \varphi(z_m) \varphi(1) \rangle_\mathrm{c}$ against $\log |z-1|$
for $\lambda=0.5,0.6,0.7$, 
where the data for $\lambda = 0.6, 0.7$ are simultaneously shifted in the vertical direction 
by $\alpha_{0.6 \to 0.5}=\log[Z(\lambda=0.5)/Z(\lambda=0.6)]=-0.015(1)$ and $\alpha_{0.7 \to 0.5}=-0.056(1)$, respectively, so that they agree with the data for $\lambda=0.5$.
In Fig.\ref{4pt_after_renormalization_oncritical},
we also plot $\log \langle \varphi(z_m) \varphi(1) \varphi(e^{i\frac{\pi}{3}})\varphi(e^{i\frac{5\pi}{3}})\rangle_\mathrm{c}$ against $\log |z-1|$
for $\lambda=0.5,0.6,0.7$, 
where the data for $\lambda = 0.6, 0.7$ are simultaneously shifted in the vertical direction 
by $2\alpha_{0.6 \to 0.5}$ and $2\alpha_{0.7 \to 0.5}$, respectively, 
so that they agree with the data for $\lambda=0.5$.
These shifts correspond to a wave function renormalization.
Namely, 2-point and 4-point correlation functions agree after performing wave function renormalization.
Therefore, the above results indicates 
that the theories are universal on the phase boundary as in ordinary field theories.
We do not see the above agreement of the correlation functions 
in either the UV region with $m=1,2$, or the IR region with $m=14, 15$.
We consider the disagreement in the latter region to be caused by increase of the ambiguity of 
the position on the fuzzy sphere due to the stereographic projection for large $|z|$.

Finally, we see a connection of the theory we have studied to a CFT. 
In Fig.\ref{2pt_after_renormalization_oncritical}, we fit seven data points ($m=4,\ldots,10$) of  
$\log \langle \varphi(z_m)\varphi(1)\rangle_\mathrm{c}$ for $\lambda = 0.5$ 
to $-u\log |z-1|+v$ and obtain $u=0.149(2)$ and $v=1.887(1)$. 
This means that the 2-point correlation function behaves as
\begin{equation}
\label{2pt function on the boundary}
\langle \varphi(z)\varphi(1)\rangle_\mathrm{c} = \frac{e^v}{|z-1|^u} \quad \mathrm{for}\ \  m=4,\ldots,10 \ .
\end{equation}
In CFTs, the 2-point correlation function of the operator $\mathcal{O}(z)$ with the scaling dimension $\Delta$ behaves as
$
\left \langle \mathcal{O}(z) \mathcal{O}(z') \right \rangle \sim 1/|z-z'|^{2\Delta}
$.
Thus, the present theory on the phase boundary behaves as a CFT in the UV region.
Our 2-point correlation function deviates universally from that in the CFT
in the IR region with $11 \leq m \leq 13$.
Moreover, in a further UV region with $m=3$, it also deviates universally.
We consider these deviations to be an effect of the UV/IR mixing.
It is nontrivial that we observe the behavior of the CFT because field theories on non-commutative spaces are non-local ones.
%
%
\section{Conclusion and discussion}
\setcounter{equation}{0}
We have studied renormalization in the scalar filed theory on the fuzzy sphere by Monte Carlo simulation.
We examined the 2-point and 4-point correlation functions on the phase boundary 
where the spontaneous $Z_2$ symmetry breaking occurs.
We found the agreement of correlation functions at different points on the boundary 
up to the wave function renormalization. 
This implies that the critical theory is universal, which is consistent with the universality
in the disordered phase \cite{Hatakeyama:2017fao, Hatakeyama:2018qjr}, 
because the phase boundary is obtained by one parameter fine-tuning. 
Moreover, it was observed that the behavior of the 2-point correlation functions is the same as that in a CFT at short distances and universally different from that at long distances.
We consider the latter to be due to the UV/IR mixing.

The CFT observed at short distances seems to be different from the critical Ising model, 
because the value of $u/2$ in \eqref{2pt function on the boundary} disagrees with the scaling dimension of the spin operator, $\Delta_\mathrm{Ising}=1/8$.
This indicates that the universality classes of the scalar field theory on the fuzzy sphere are totally different from those of an ordinary field theory\footnote{It should be noted that $\Delta_\mathrm{ours} = u/2 \simeq 0.075 = 3/40$ coincides with the scaling dimension of the spin operator in the tricritical Ising model, which is the $(4, 5)$ unitary minimal model.}.

In fact, the authors of \cite{Martin:2004un, Panero:2006bx, Panero:2006cs, GarciaFlores:2009hf, Das:2007gm} reported that there exists a novel phase in the theory on the fuzzy sphere that is called the non-uniformly ordered phase \cite{Gubser:2000cd,Ambjorn:2002nj}.
We hope to clarify the universality classes by studying renormalization in the whole phase diagram.

\acknowledgments
Numerical computation was carried out on XC40 at YITP in Kyoto University and FX10 at the University of Tokyo.
The work of A.T. is supported in part by Grant-in-Aid
for Scientific Research
(No. 15K05046)
from JSPS.

\end{document}